\documentclass[12pt,aps,amsmath,amssymb]{article}
\pdfoutput=1
\usepackage{graphicx,epsf}
\usepackage[usenames]{color}
\usepackage[]{latexsym}
\usepackage{bbm}
\usepackage{dsfont}
\usepackage[utf8]{inputenc}

\newcommand{\be}{\begin{eqnarray}}
\newcommand{\ee}{\end{eqnarray}}
\renewcommand{\d}{\mbox{{\rm d}}}
\definecolor{yellow}{rgb}{0.95,0.75,0.1}
\definecolor{red}{rgb}{1,0,0}
\definecolor{green}{rgb}{0,1,0}
\definecolor{blue}{rgb}{0,0.5,1}

\definecolor{lblue}{rgb}{0,0.8,1}
\definecolor{dblue}{rgb}{0,0,1}
\definecolor{dgreen}{rgb}{0,0.5,0}
\definecolor{lila}{rgb}{0.8,0,0.8}
\definecolor{violet}{rgb}{0.25,0,1}
\definecolor{grey}{rgb}{0.3,0.3,0.3}

\definecolor{contoura}{rgb}{0,0,1}
\definecolor{contourb}{rgb}{0,1,1}
\definecolor{contourc}{rgb}{0,1,0}
\definecolor{contourd}{rgb}{0.95,0.75,0.1}
\definecolor{contoure}{rgb}{1,0,0}
\definecolor{contourf}{rgb}{1,0,1}

\title{Rotating AdS Einstein universes from constrained metrics}
\author{Benjamin Harms\thanks{Email: bharms$@$ua.edu}
\\ \small{Department of Physics \& Astronomy, The University of Alabama}
\\ \small{Tuscaloosa, Alabama 35487-0324, USA}}
\date{\today}

\begin{document}
\maketitle

\abstract{The imposition of a constraint between the metric tensor elements in both  three- and four-dimensional, rotating AdS space-times is shown to reduce the number of independent equations of motion and to result in new families of solutions to the equations of motion.  For the geometries investigated analytic solutions or partial analytic solutions of the equations of motion are obtained.  In all cases the number of independent field equations is less than the number of independent functions, resulting in an undetermined function which can be freely specified. For rotating, asymptotically $AdS$ space-times the reduction of the number of field equations to be solved holds for vacuum black hole solutions and for black hole solutions obtained from space-times containing matter.   }

\section{Introduction}

In two recent works,\cite{Harms:2016pow},\cite{Harms:2017yko} novel classical solutions to  gravity coupled to matter were obtained in an asymptotically $AdS$ background, which represent topological solitons and hairy black holes.
The requirement that the space-time be asymptotically $AdS_3$ led to the discovery of a hidden local symmetry of the system which entangles the space-time metric with the matter fields.    Equivalence classes of hairy black hole solutions were obtained for the resulting space-times.

 Asymptotically $AdS$  solutions to Einstein equations are of  current interest due to their application to holography and their possible indication of phase transitions in the boundary field theory~\cite{Witten:1998zw}. Known solutions to  gravity with asymptotically $AdS$ boundary conditions are the $AdS$ black hole~\cite{Banados:1992gq} and the $AdS$ soliton~\cite{Horowitz:1998ha}.
The $AdS$ black hole is thought to correspond to a plasma, or deconfining phase, in the boundary theory, while the $AdS$ soliton corresponds to a confining phase.  By comparing their free energies, it was determined that the $AdS$ soliton, describing the confining phase, is stable at low temperatures, and the $AdS$ black hole, describing the plasma phase, is stable at high temperatures.  The phase  transition was determined to be first order. (These issues are reviewed in \cite{Natsuume:2014sfa}.)  

In addition to the  $AdS$ black hole and  $AdS$ soliton,  a number of other solutions have been found to the Einstein equations, with gravity minimally coupled to matter fields, which are also  asymptotically $AdS$.  This indicates the possibility of a richer phase structure in the dual theory.  Solutions of this type may or may not have horizons, and have been classified as  hairy versions of the $AdS$ black hole or the  $AdS$ soliton, respectively.\cite{Henneaux:2002wm,Banados:2005hm,Brihaye:2013tra,Anabalon:2016izw}.  In  \cite{Harms:2016pow},\cite{Harms:2017yko} 
new types of such asymptotically $AdS $ solutions  in three space-time dimensions were found.  Their matter contribution is the nonlinear  $\sigma$-model, and they describe both self-gravitating topological solitons and hairy black holes .

 The existence of static, self-gravitating, nonlinear $\sigma$-model solitons in an asymptotically flat  $2+1 $ dimensional space-time has been known  for a long time~\cite{Clement:1976hh}.  On the other hand, it was previously shown  that static (non-rotating) nonlinear $\sigma$-model solitons which are asymptotically $AdS_3 $ do not exist.\cite{Bizon:2004wa}    This is also evident  from a simple scaling argument. While the  standard $\sigma$-model Lagrangian, consisting of only a quadratic term, evaluated for static field configurations, is scale invariant in two spatial dimensions, this is no longer the case in a background anti-de Sitter space.  For the latter, the cosmological constant contribution scales like the square of the radial distance leading to an attractive force,  which causes the collapse of any nonrotating field configurations.  In \cite{Harms:2016pow}  stability was shown to be recovered for rotating field configurations, and as a result  topological solitons can exist in a space which is asymptotically $AdS_3 $.  Numerical solutions were obtained for the asymptotically $AdS_3$ topological solitons, along with their masses and moments of inertia in a collective coordinate approximation. Upon embedding the solutions in 3 + 1 dimensions, they can be interpreted as cosmic strings.
 No  horizons appeared for any of these solitons, and therefore they do not correspond to black holes.

In  \cite{Harms:2017yko}  only the quartic  term in the standard nonlinear $\sigma-$model Lagrangian was kept.
 After coupling this system to gravity in an asymptotically  $AdS$ space-time, the system
was shown to not possess  soliton solutions, but instead infinitely many hairy black hole solutions.  This is due to  the presence of a novel local symmetry which arises from the imposition of a constraint on a subset of the metric tensor elements for this self-gravitating non-linear $\sigma-$model.  The gauge  transformation simultaneously twists the sigma model fields and changes the space-time metric.   Using this gauge symmetry,  an  extremal $BTZ$ black hole can be continuously transformed to infinitely many hairy black hole field configurations, all of which are solutions to the coupled  Einstein and matter field equations.  The hairy black hole solutions have an asymptotic behavior which  resembles that of extremal black holes, and they are referred to as  extremal hairy black holes.
Previous searches for solutions for $AdS$  
black holes with scalar hair  generally required introducing  complicated expressions for the potential energy density. In \cite{Harms:2017yko} the existence of hairy black hole  solutions was shown to be  due primarily to the existence of the hidden gauge  symmetry.
Unlike the hairy black holes found in other models, in the quartic-only model  the energy-momentum source is  light-like, and the Hawking temperature is zero.  The action of the local symmetry transformation leaves these results unchanged.

\section{Vacuum  black hole solutions in three and four dimensions }
\subsection{Three-dimensional dilatonic black holes}
The imposition of a metric constraint to obtain a family of black hole solutions applies to other three-dimensional black holes as well as the family of hairy black holes described above.  As an example, the three-dimensional dilatonic black hole described in \cite{Lemos:1994xp} is characterized after appropriate normalization by an action of the form
\be
S\,=\,\frac{1}{2\pi} \int\,d^3\,x\sqrt{-g}\,{\rm e}^{-2\,\Phi(r)}\,\left(R-2\,\Lambda\right)\, ,
\ee
where $\Phi(r)$ is the dilaton field, $R$ is the Ricci scalar and $\Lambda$ is the cosmological constant.  The general form of the metric is 
\be 
ds^2\,=\,-A(r)\,dt^2+B(r)\,dr^2+2\,H(r)\,dt\,d\phi + M(r)\,d\phi^2\, ,
\label{3metric}
\ee
and a family of three-dimensional black hole solutions exists for the constraint
\be 
A(r)\,=\, \omega^2\,M(r) + 2\,\omega\,H(r)\, .
\label{cons}
\ee
In this relation $\omega$ is an arbitrary constant.
For this constraint the $\sqrt{-g}$ factor in the action simplifies to the expression
\be 
\sqrt{-g}\,=\,\sqrt{B(r)}\,G(r)\, ,
\ee
in which the function $G(r)$ is defined as $G(r)\,=\,\omega\,M(r)+H(r)$. As was true for the field equations obtained for the hairy black holes, the constraint in Eq.(\ref{cons}) reduces the number of field equations to one less than the number of unknown functions, leaving one function undetermined.  The metric tensor elements in Eq.(\ref{3metric}) can all be expressed in terms of the arbitrary function $G(r)$.  Analytic expressions in terms of $G(r)$ for the fields $B(r),\, H(r),$ and $\Phi(r)$ can be obtained from the field equations, allowing $A(r)$ and $M(r)$ to be written in terms of $G(r)$ as well,
\be 
A(r)&=&\omega\, G(r)\,(1+W(r))\nonumber \\
B(r)&=& \frac{3\,G'(r)^2}{\sqrt{G(r)}\left(3\,b_1 -4\,\Lambda\,G(r)^{3/2}\right)} \nonumber\\
H(r)&=& G(r)\,W(r)\nonumber\\
M(r)&=&G(r)\,(1-W(r))/\omega\, .
\label{Gmetric}
\ee
In these equations $W(r)$ is  given by
\be 
W(r)\,=\,c_1+\frac{c_2}{9\, b_1}\left(\ln\left[\frac{G(r)^3}{(3\,b_1-4\,\Lambda G(r)^{3/2})^2}\right]\right)\, ,
\label{WFN}
\ee
where the prime indicates a derivative with respect to $r$, and $b_1\, ,c_1\, ,c_2\, ,$ and $\Lambda\,(<\,0)$ are constants.  An event horizon exists for $G(r_H)\,=\,(3\,b_1\,/(4\,\Lambda))^{2/3}$ for $b_1\,<\,0$.  The dilaton field is 
\be 
{\rm e}^{-2\,\Phi(r)}\,=\,c\,\alpha\,\left[\frac{G(r)^{1/4}}{(3\,b_1-4\,\Lambda\,G(r)^{3/2})^2}\right]\, ,
\ee
 where $c$ is an arbitrary constant, and $\alpha\,=\,\sqrt{-\Lambda/3}$.
\paragraph{}
The masses and angular momenta of the black holes obtained from the geometries described by Eq.(\ref{Gmetric}) can be determined by writing the action in the Hamiltonian formulation \cite{Arnowitt:1962hi,Hanson:1976cn}
\be 
S\,=\,\int\,H\,dt + B_b\, .
\label{Haction}
\ee
In this expression $B_b$ is a boundary term and 
\be 
H\,=\,\int\,d^2 x \left[ N_0\,{\cal{H}}_0+N_\phi\,{\mathcal{H_\phi}}\right]\, ,
\ee
with
\be 
\mathcal{H}_0&=&\gamma^{-1/2}\left[\pi^{i j}\,\pi_{i j} - \pi^2\right] -\gamma^{1/2}\left(R- 2\,\Lambda\right)\nonumber\\
\mathcal{H_\phi}&=&-\,2\,{{\pi_i^{i j}}_|}_j\, .
\ee
In these expressions $\gamma$ is the determinant of the spatial part of the metric, $N_0$ and $N_\phi$ are the lapse and shift functions respectively. The $\pi^{i j}$'s are the momenta conjugate to the metric tensor elements, and they are determined by the extrinsic curvature. 
\paragraph{} 
For the dilatonic black hole the metric is written as 
\be 
ds^2\,=\, -N_0(\rho)\, dt^2+\rho^2\,\left(N_\phi(\rho)\, dt+\phi\right)^2+\frac{d\rho^2}{f(\rho)^2}\, .
\label{lsmetric}
\ee
The metric tensor elements in this expression are related to those in the  metric in Eq.(\ref{3metric})  by the relations
\be 
\rho^2&=&M(r)\nonumber\\
\rho^2\,N_\phi(\rho)&=&H(r)\nonumber\\
N_0(\rho)^2-\rho^2\,N_\phi(\rho)^2&=&A(r)\nonumber\\
f(\rho)^2&=&\left(B(r)\right)^{-1}\,\left(\frac{d\rho}{dr}\right)^2\, .
\ee
The action as written in Eq.(\ref{Haction}) for the metric in Eq.(\ref{lsmetric}) is varied with respect to $f^2,\, \pi^{i j}$ and $\Phi(\rho)$.  For the three-dimensional black hole the only non-zero $\pi^{i j}$ is $\pi^{r \phi}\,=\,\rho^3\,N_{\phi}(\rho),_{\rho}\,/\,N(\rho)$, where $N(\rho)\,=\,N_0(\rho)\,/f(\rho)$ \cite{Banados:1992gq,Lemos:1994xp}.  The mass and angular momentum can be obtained from the surface terms which arise in the variation process.  The variations of the action are rendered finite by the usual procedure of the subtraction of the AdS background.
\paragraph{}
To obtain explicit expressions for the mass and the angular momentum the function $G(r)$ must be specified.  In order for the metric to be asymptotically AdS, $G(r)$ could, for example, be chosen to be $G(r)\,=\,r^2$.  For this simple case the variation of the surface action terms leads to values of the mass and angular momentum which are somewhat complicated due to the number of undetermined constants in the expressions in Eq.(\ref{Gmetric}) and Eq.(\ref{WFN}) for the metric tensor elements and the dilaton field. For the parameter choices $\Lambda\,=\,-3,\, \alpha\,=\,1,\, c_1\,=1$  
\be 
M_{BH}&=&\frac{-3\,b_1\,\left(47-39\,\omega\right)}{16\,\ln(12)}\nonumber\\
& &\\
J_{BH}&=&\frac{3}{64\,\omega^{3/2}}\left(27 -\frac{3\,\left(9\,\omega-4(\ln(12))^2/b_1\right)}{2\ln(12)}\right)^{1/2}\, ,\nonumber
\ee
for the black hole mass and angular momentum respectively.
\subsection{Four-dimensional black holes}
The most general form of the Petrov type-D solution of the source-free Einstein-Maxwell equations was obtained by Plebanski and Demianski.  The 
  Pleba\'nski and Demia\'nski  metric,\cite{Plebanski:1976gy} can be 
expressed in coordinates $\{t,r,p,\sigma \}$, where $t\in R$,  $r\in R_+, $  $-a\le p\le a$ (for some $0<a<\infty$) and $0\le\sigma<2\pi$.  After a re-scaling,\cite{Klemm:1997ea} the metric tensor can be written as
\be 
ds^2\,=\,\frac{-Q(r)(d\tau-p^2 d\sigma)^2}{\rho^2}+\frac{P(p)(d\tau+r^2 d\sigma)^2}{\rho^2}+\frac{\rho^2}{P(p)}dp^2+\frac{\rho^2}{Q(r)}dr^2\, ,
\label{pdmetric}
\ee
where $\rho^2\,=\,r^2+p^2$ and
\be 
Q(r)&=&\alpha+q_e^2+q_m^2-2 m r +\epsilon r^2 -2 n r^3 - (\Lambda/3)\, r^4\nonumber \\
\\
P(p)&=&\alpha+2 n p -\epsilon p^2 +2 m p^3 - (q_e^2+q_m^2+\Lambda/3)\, p^4\, .\nonumber
\label{PQ}
\ee
In these equations $q_e$ and $q_m$ are the electric and magnetic charges, $\Lambda$ is the cosmological constant, $m$ is the mass parameter, $n$ is the NUT-charge.  The remaining parameters $\alpha$ and $\epsilon$ are arbitrary.  

The local symmetry between nonlinear model fields and the space-time discovered in the rotating 2+1 $AdS$ space-time found in \cite{Harms:2017yko} also exists in the four dimensional solution in Eq.(\ref{pdmetric}). For certain choices of the parameters in Eq.(\ref{PQ}) a constraint condition among the metric tensor elements, which is the four-dimensional counterpart of the constraint found in  \cite{Harms:2017yko}, holds.  Written in the same form as in \cite{Harms:2017yko}
\be 
-g_{\tau\tau}\,=\,\omega^2\,g_{\sigma\sigma} + 2\,\omega\,g_{t\sigma}\, .
\label{gauge}
\ee
 This symmetry requires the functions $Q(r)$ and $P(p)$ to satisfy
\be 
Q(r) - P(p)\,=\,\omega^2\,(-Q(r)\,p^4 + P(p)\,r^4) + 2\,\omega\,(Q(r)\,p^2 +r^2\,P(p))\, ,
\label{QmP}
\ee
where $\omega$ is an arbitrary constant.  This condition requires that 
\be 
Q(r)&=&(r^2+1/\omega)^2/l^2 \nonumber\\
\\
P(p)&=&(p^2-1/\omega)^2/l^2\nonumber\, ,
\ee
where $l^2\,=\,-3/\Lambda$.
These forms for $Q(r)$ and $P(p)$ are consistent with the forms in Eq.(\ref{PQ}) for the parameter sets $q_e\,=\,q_m\,=\,m\,=\,n\,=\,0$ and $\alpha\,=\,-\Lambda/(3\, \omega^2), \epsilon\,=\,-2 \Lambda/(3\, \omega )$.  The condition in Eq.(\ref{QmP}) is very restrictive in that it holds only for a non-vanishing cosmological constant, for zero mass and no coupling to matter.  Nevertheless it provides a starting point from which an investigation of 
four-dimensional, rotating space-times with matter content can begin.  For $\omega\,<\,0$ an event horizon exists at $r_H\,=\,1/\sqrt{-\omega}$.  To compare the geometry which results from the constraint imposed in Eq.(\ref{gauge}) to the geometry discovered in \cite{Klemm:1997ea} the coordinates can be transformed to 
\be 
d\tau\,=\,\Xi\, \frac{dt - a\,d\phi}{\Xi}\, ,\hspace{5mm}d\sigma\,=\,-d\phi/(a\,\Xi)\, ,\hspace{5mm} p\,=\,a\cosh(\theta)\, .
\ee
 In these coordinates the invariant line element becomes
\be 
ds^2&=&\rho^2\left(\frac{dr^2}{Q(r)}+ \frac{l^2 a^2\sinh(\theta)^2\,d\theta^2}{(a^2\cosh(\theta)^2 - 1/\omega)^2}\right)+\frac{\left(a^2\cosh(\theta)^2-1/\omega\right)^2}{a^2 \rho^2 \Xi^2}\left[a\,dt - (r^2 + a^2)\d\phi\right]^2\nonumber\\
 &-& \frac{Q(r)}{\rho^2 \Xi^2}\left[dt + a\sinh(\theta)^2 d\phi\right]^2\, ,
 \label{hyperbh}
\ee
where $\rho^2\,=\,r^2+a^2\,\cosh(\theta)^2$ and $\Xi\,=\,1+a^2/l^2$.

For negative $\omega$ the metric in Eq.(\ref{hyperbh}) describes a rotating black hole with a hyperbolic horizon surface very similar to but distinct from the one in \cite{Caldarelli:1998hg}.  

\section{Solutions for four-dimensional space-times with matter}
  A natural candidate for the inclusion of matter in four space-time dimensions would be the  Skyrme model coupled to gravity.  The Skyrme model coupled to gravity in $3+1$ space-time has been studied for some time and shown to admit solitons and  hairy black holes.\cite{Heusler:1991xx}-\cite{Gudnason:2016kuu} The quadratic term in the chiral fields was present in these articles. The analysis in \cite{Harms:2017yko} shows  that a hidden local symmetry exists between  chiral   fields and the space-time metric if the quadratic term is absent, and the only matter contribution to the action is the quartic (Skyrme) term.

The Lagrangian density for the gravitating Skyrme model with only the quartic matter contribution is 
\be 
{\mathcal{L}}\,=\, \sqrt{-g}\,\left(\frac{R}{16\pi G} + \frac{1}{32 e^2}Tr\left(\left[ K_{\mu}, K_{\nu}\right] \left[ K^{\mu}, K^{\nu}\right]\right)\right) \, , 
\ee
where $G$ is the gravitational constant, $g$ is the determinant of the metric tensor, $R$ is the Ricci scalar, ${\rm K_{\mu}\,=\,\partial_{\mu} U\,U^{-1}}$, and $U$ is an $SU(2)$-valued field in the defining representation.  A field rotating around the $z-$direction in internal space with angular velocity $\omega$ is described by
\be
U \,=\, n_1(r,p)  \mathds{1} + i n_3(r,p) \tau_z + i n_2(r,p) \left( \tau_x \cos(\varphi - \omega t) + \tau_y \sin(\varphi -\omega t) \right)\, , 
\ee
where $\mathds{1} $ is the $2\times 2$ unit matrix and $\tau_{x,y,z}$ are the Pauli matrices.  $\vec n=(n_1,n_2,n_3)$ is a unit vector.
The $n_i$ are restricted to being functions of only a radial coordinate $r$ and a certain projection along a constant direction, $p\,=\, a\,\cos(\theta)$, where $a$ is an arbitrary constant which can be interpreted as a rotation parameter.  $\phi$ and $t$ are angular and time coordinates, respectively.
The general form for the  invariant measure describing  space-time rotation which is rotationally invariant  about one axis involves five functions of $r$ and $p$, which shall be denoted by $A,\, B,\,F,\,H$ and $M$, 
The metric is taken to be
\be 
ds^2\,=\,-A(r,p)\, dt^2 + B(r,p)\, dr^2 + F(r,p)\,\,dp^2+ H(r,p)\,dt d\phi + M(r,p)\,d\phi^2\, . 
\label{metric}
\ee
A simple choice for the $n_i$'s is  $n_3\,=\,0$, and
\be 
n_1(r,p)\,=\,\cos(\chi(r,p))\, ,\hspace{1cm} n_2(r,p)\,=\, \sin(\chi(r,p))\, .
\ee
This is one of the choices for $n_1, n_2$ and $n_3$ analyzed in  \cite{Ioannidou:2006nn}.  For these choices of $n_1\, ,n_2$ and $n_3$, the trace of the product of the commutators is
\be 
Tr\left(\left[ K_{\mu}, K_{\nu}\right] \left[ K^{\mu}, K^{\nu}\right]\right)= W(r,p)\left(\frac{1}{B(r,p)}\left(\frac{\partial\chi(r,p)}{\partial r}\right)^2  + \frac{1}{F(r,p)}\left(\frac{\partial\chi(r,p)}{\partial p}\right)^2 \right).
\label{Tr}
\ee
where
\hspace{-2cm} 
{ \be 
 W(r,p) = 8\,{\frac { \left( \sin \left( \chi \left( r,p \right)  \right) 
 \right) ^{2}\,\sqrt{-g} \left( -A
 \left( r,p \right) +M\left( r,p \right) {\omega}^{2}+2\,H
 \left( r,p \right) \omega \right) }{\left( A\left( r,p \right) M\left( r,p \right) + \left( H\left( r,p \right)  \right) ^{2} \right) }}
 \label{Wrp}
\ee}
The right hand side is zero for the constraint in Eq.(\ref{gauge}),
\be 
 A(r,p)\,=\,M \left( r,p \right) {\omega}^{2}+2\,H\left( r,p \right) \omega \, .
 \label{Arp}
 \ee 
 This is the counterpart in this model to the expression found for $A(r)$ for the spinning Skyrmion in $AdS_3$ \cite{Harms:2017yko}.
 \par
 Before replacing $A(r,p)$ by the expression in terms of $M(r,p)$ and $H(r,p)$, the field equations are calculated.  The result is a set of six equations, two of which are independent of $\chi(r,p)$. One of the four remaining equations can be used to solve for $\chi^{(1,0)}(r,p)$, which is then replaced in the remaining three equations.  The expression for $A$ in terms of $M$ and $H$ reduces the sixth equation, which is obtained from the variation $\delta\mathcal{L}/\delta\chi$, to zero, leaving four coupled equations in terms of the four unknown functions $B, F, H$ and $M$.  The substitution $M(r,p)\,=\,(G(r,p) - H(r,p))/\omega$ makes two of the equations equivalent, leaving three equations in the three unknowns $B(r,p),\, F(r,p),$ and $G(r,p)$. These three functions can be determined analytically (independently of the specification of the scalar function $\chi(r,p)$) 
\be
B(r,p)&=&\frac{\rho(r,p)^2}{(\alpha r^2 - \beta)^2}\, ,\nonumber\\
F(r,p)&=&\frac{\rho (r,p )^2}{(\alpha p^2 + \beta)^2}\, ,\\
G(r,p)&=&\left(\frac{4\,C_1 r^2 + \Gamma)}{\Gamma_1}\right)\,\left(\frac{4\,C_1 p^2 - \Gamma}{\Gamma_2}\right)\, , \nonumber
\label{BFG}
\ee
where  $\rho(r,p)^2\,=\,p^2+r^2$; $\Gamma$, $\Gamma_1$, and $\Gamma_2$ are  the constants of integration; and $\alpha$ and $\beta$ are constants which depend upon the cosmological constant and $\Gamma$,
 \be 
\alpha\,=\,\frac{-2 C_1}{3}, \, \beta\,=\,\frac{\Gamma}{12 C_1}, \,C_1\,=\,\pm\sqrt{-\frac{3 \Lambda}{4}}\, .
 \ee
A horizon exists at $r_H\,=\,\sqrt{\beta/\alpha}\,=\,\sqrt{\Gamma/6\,\Lambda}$, for $\Gamma\,<\,0$.

\begin{figure}[ht]
\center
\includegraphics[height=6.1cm]{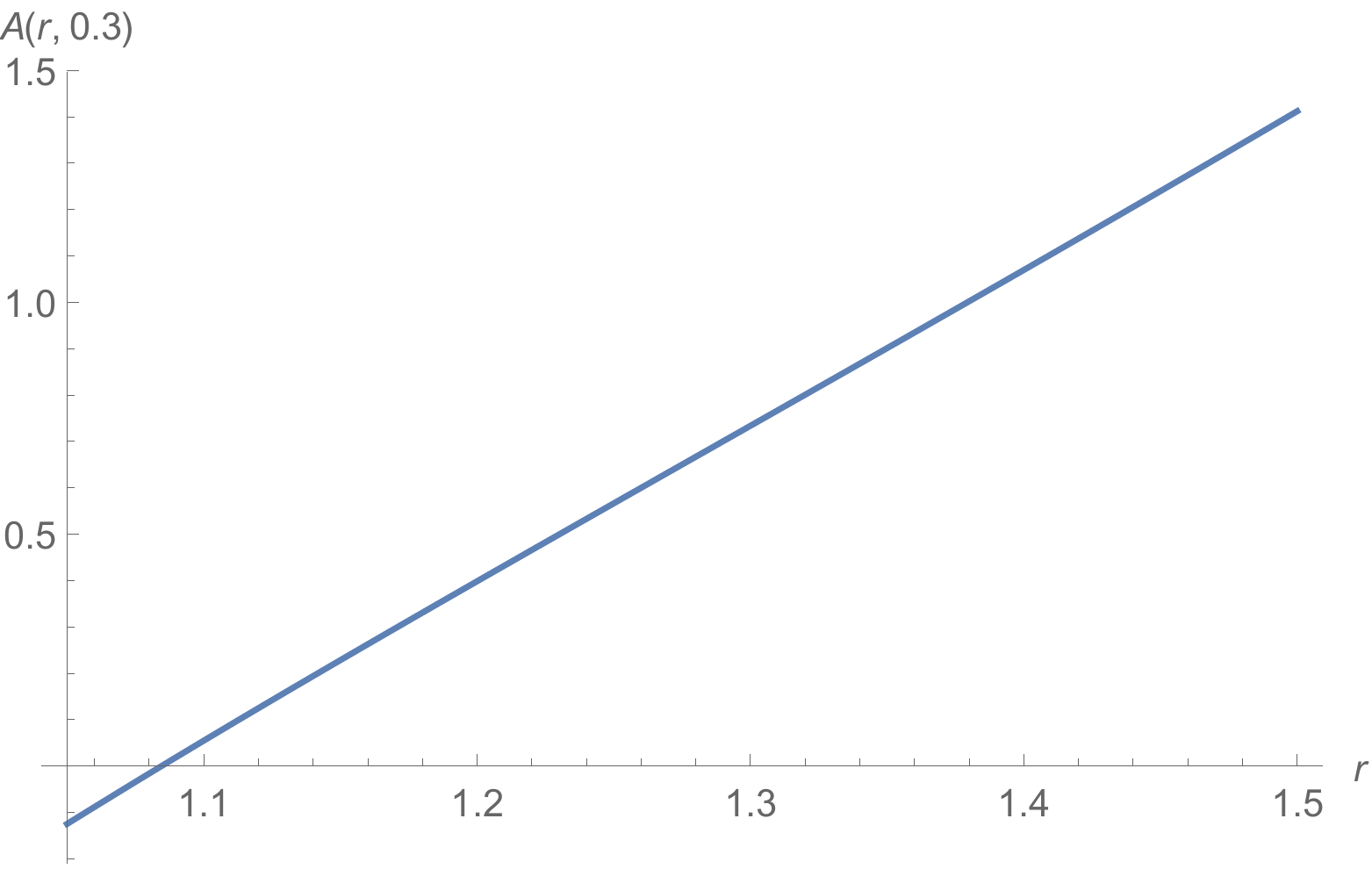}
\caption{\footnotesize $A(r,p)$ as a function of $r$ for $p = 0.3,\, \alpha\,=\,1,\, \beta\,=\,1/9,\, \kappa\,=\,0.01,\,  \omega\,=\,-1$.  $A(r,p)$ vanishes at $r\,=\,1.08$.  As $r$ approaches infinity, $ A(r,p) \to r^2$. }
\label{Avsr}
\end{figure}

To obtain the remaining functions $A(r,p), H(r,p), M(r,p), \chi(r,p)$ one of them must be specified.  The other three can then be determined from the relations given in Eqs.(\ref{Wrp}-\ref{BFG}) or from the equations of motion.  As an example, the choice 
\be 
\chi(r,p)\,=\,2\,\arctan(\frac{a}{r})\, ,
\ee
  where $a$ is an arbitrary constant, allows $H(r,p)$ to be obtained by numerically solving   the field equation which contains $B(r,p),\, F(r,p)\, , G(r,p)\, , \chi(r)\,$ and $H(r,p)$.  For the parameters given in Fig.\ref{Avsr} the event horizon is at $r_{eh}\,=\,0.11$ and the surface of infinite redshift, $A(r_{rs},p)\,=\,0$, is at $r_{rs}\,=\,1.08$. $A(r,p)$ is obtained from the relation $A(r,p)\,=\, \omega\, (G(r,p)+H(r,p))$.  In the asymptotic limit, $r\to \infty\, ,\, A(r,0.3) \to r^2$. 
\section{Discussion}

The study of space-times which are asymptotically $AdS_4$ obtained via the procedure described above is of particular interest due to the AdS/CFT correspondence.  One goal will be to determine the effect of this hidden symmetry on the conformal field theories which live on the two- and three-dimensional boundaries of the three- and four-dimensional space-times which can be generated by various choices of the undetermined fields, e.g. the scalar field $\chi(r,p)$ in the model discussed in Section {\bf 3}.

 The vanishing of the matter contribution to the on-shell action for field configurations satisfying the equation for $A(r,p)$ given above
occurs in five dimensions, just as it does in four dimensions.  A search for solutions of the field equations for the five dimensional case have not yet been carried out, but the general form of the field equations is the same as in four dimensions.  The same techniques used in the four-dimensional case can be used to attempt to find analytic solutions in five dimensions.  If analytic solutions are not possible, numerical methods will be used to obtain solutions.  The solutions obtained in this case will allow the effect of the constraint on the four-dimensional conformal field theories on the boundary to be investigated. Applying the method used in the three- and four-dimensional models, the number of field equations to be solved in the five-dimensional case can be reduced from 7 to 4.  This will facilitate the search for solutions and the analysis of the effect of the constraint on the conformal field theories living on the four-dimensional boundary.

 Another interesting line of research is the study of the dynamical stability of the solutions obtained in four- and five-dimensional versions of the Skyrme model coupled to gravity in a rotating universe.  In \cite{Heusler:1991xx} the linear stability of self-gravitating Skyrmions against small, time-dependent perturbations was investigated for the particle-like solutions of the Einstein-Skyrme equations on a static background.  An investigation of the stability of the particle-like solutions and the black hole solutions on a rotating background which is asymptotically $AdS_n$ is another important project.

 Since the vanishing of the matter content of the on-shell Lagrangian density for restricted field configurations occurs in 3, 4, and 5 dimensions, it is reasonable to assume that this result holds in any number of dimensions.  Thus the effect of constraints on the matter content of the Lagrangian density in an arbitrary number of dimensions and the solutions to the corresponding field equations is another interesting topic of investigation.
 
The simplification of the systems of equations which arises due to imposing the constraint
\be 
-g_{t t}\,=\,\omega^2\,g_{\phi \phi} + 2\,\omega\,g_{t \phi}
\label{mcon}
\ee
has been exploited only for rotating space-times which are asymptotically $AdS$.  The constraint in Eq.(\ref{mcon}) may be useful only for such space-times.  However, other constraint equations may reveal simplifications in space-times which are not rotating, asymptotically $AdS$.  The challenge is to find the principle which governs the choice of the simplifying equation of constraint.
 
\section*{Acknowledgements}
The author is grateful to A. Stern, C. Cartwright and M. Kaminski for useful discussions and technical assistance.

\end{document}